\documentclass[useAMS,usenatbib]{mn2e}

\usepackage{graphicx}
\usepackage{graphicx}
\usepackage{savesym}
\usepackage{amsmath}
\savesymbol{iint}
\usepackage{txfonts}
\restoresymbol{TXF}{iint}
\usepackage{amssymb}
\usepackage{epsfig}
\usepackage{natbib}
\usepackage{multirow}
\usepackage{hyperref}

\newcommand{\xmm}{\emph{XMM-Newton}}

\newcommand{\spitzer}{\emph{Spitzer}}
\newcommand{\herschel}{\emph{Herschel}}

\newcommand{\lumunit}{erg~s$^{-1}$}
\newcommand{\fluxunit}{erg~s$^{-1}$~cm$^{-2}$}

\newcommand{\msun}{$M_\odot$}
\newcommand{\Lsun}{$L_\odot$}

\newcommand{\slope}{$\Gamma_{\rm X}$}

\newcommand{\avg}[1]{\left< #1 \right>}

\title[The starburst-AGN connection in the merger galaxy Mrk 938]{The starburst-AGN connection in the merger galaxy Mrk 938: \\ an infrared and X-ray view\thanks{{\it Herschel} is an ESA space observatory with science instruments provided by European-led Principal Investigator consortia and with important participation from NASA.}}
\author[P. Esquej et al.]
{\parbox{\textwidth}{P. Esquej$^{1,2,3}$\thanks{E-mail: pilar.esquej@cab.inta-csic.es}, 
A. Alonso-Herrero$^{1,2}$\thanks{Augusto Gonz\'alez Linares Senior Research Fellow}, 
A.M. P\'{e}rez-Garc\'{\i}a$^{4,5}$,
M. Pereira-Santaella$^{1}$, 
D. Rigopoulou$^{6}$, 
M. S\'{a}nchez-Portal$^{7}$,
M. Castillo$^{7}$,
C. Ramos Almeida$^{8}$,
D. Coia$^{7}$,
B. Altieri$^{7}$,
J.A. Acosta-Pulido$^{4,5}$,
L. Conversi$^{7}$,
J.I. Gonz\'{a}lez-Serrano$^{2}$,
E. Hatziminaoglou$^{9}$,
M. Povi\'{c}$^{10}$,
J.M. Rodr\'{\i}guez-Espinosa$^{4,5}$,
I. Valtchanov$^{7}$
}\vspace{0.4cm}\\
$^{1}$Centro de Astrobiolog\'ia (CSIC--INTA), Dept. Astrof\'isica, ESAC Campus, P.O. Box 78, 28691 Villanueva de la Ca\~nada, Spain\\
$^{2}$Instituto de F\'{\i}sica de Cantabria, CSIC-Universidad de Cantabria, 39005 Santander, Spain\\
$^{3}$ Departamento de F\'{\i}sica Moderna, Universidad de Cantabria, Avda. de Los Castros s/n, 39005 Santander, Spain\\
$^{4}$Instituto de Astrof\'{\i}sica de Canarias (IAC), C/V\'{\i}a L\'{a}ctea, s/n, E-38205, La Laguna, Tenerife, Spain\\
$^{5}$Departamento de Astrof\'{\i}sica, Universidad de La Laguna, E-38205 Tenerife, Spain\\
$^{6}$Astrophysics Department, University of Oxford, Oxford OX1 3RH, United Kingdom\\
$^{7}$Herschel Science Centre, INSA/ESAC, Madrid, Spain\\
$^{8}$Department of Physics and Astronomy, University of Sheffield, Sheffield, S3 7RH, UK\\
$^{9}$European Southern Observatory, Karl-Schwarzschild-Str. 2, 85748 Garching bei M\"{u}nchen, Germany\\
$^{10}$Instituto de Astrof\'{\i}sica de Andaluc\'{\i}a (CSIC), Apdo. 3004, 18080, Granada, Spain}
\begin{document}

\date{Accepted 2012 February 20. Received 2012 January 9; in original form 2011 September 16}

\pagerange{\pageref{firstpage}--\pageref{lastpage}} \pubyear{2011}

\maketitle

\label{firstpage}

\begin{abstract}
Mrk~938 is a luminous infrared galaxy in the local Universe believed to be the remnant of a galaxy merger. It shows a Seyfert 2 nucleus and intense star formation according to optical spectroscopic observations. We have studied this galaxy using new \herschel\ far-IR imaging data in addition to archival X-ray, UV, optical, near-IR and mid-IR data. Mid- and far-IR data are crucial to characterise the starburst contribution, allowing us to shed new light on its nature and to study the coexistence of AGN and starburst activity in the local Universe. The decomposition of the mid-IR \spitzer\ spectrum shows that the AGN bolometric contribution to the mid-IR and total infrared luminosity is small ($L_{\rm bol}(AGN)/L_{\rm IR}\sim$0.02), which agrees with previous estimations. We have characterised the physical nature of its strong infrared emission and constrained it to a relatively compact emitting region of $\leq2$\,kpc. It is in this obscured region where most of the current star formation activity is taking place as expected for LIRGs. We have used \herschel\ imaging data for the first time to constrain the cold dust emission with unprecedented accuracy. We have fitted the integrated far-IR spectral energy distribution and derived the properties of the dust, obtaining a dust mass of $3\times10^{7}\,{\rm M}_\odot$. The far-IR is dominated by emission at 35\,K, consistent with dust heated by the on-going star formation activity.
\end{abstract}


\section{Introduction}\label{sec:sec1}

Observations over the past decade have revealed that supermassive black holes (SMBHs) likely reside at the centres of all galaxies with spheroids \citep[see][for a review]{Kormendy1995} and that the properties of these black holes and their host galaxies are tightly correlated \citep[e.g.][]{Marconi2003, Ferrarese2005}. In addition, the similarity in the anti-hierarchical evolution of both star formation rate and active galactic nuclei (AGN) activity at z$<$2 \citep{LaFranca2005, Arnouts2007} reveals that the assembly and evolution of galaxies and the SMBHs at their centres are intimately connected. However, the nature of such an intrinsic connection is still enigmatic. Since it is believed that the main phase of SMBH growth through major mergers occurs in environments heavily obscured by dust \citep{Hopkins2006}, the infrared (IR) domain is the path to follow to unveil key signatures that remain lurking at other wavelengths. 

The class of objects known as luminous infrared galaxies (LIRGs; $L_{\rm IR}=10^{11}-10^{12}\,L_\odot$, IR in the range 8$-$1000\,\micron) contributes a significant fraction ($\geq$50\%) of the cosmic infrared background. In addition, they are major contributors to the star-formation activity at z$\sim$1 \citep{Elbaz2002, LeFloch2005, PerezGonzalez2005}. LIRGs are  powered by star formation and/or AGN activity, and a large fraction of them, especially at the high IR luminosity end, are classified as interacting galaxies and mergers \citep{Sanders1996}. It has been suggested that the mechanisms that trigger mass accretion onto the SMBH through major merger processes are episodes of nuclear star formation \citep{Hopkins2006}. Therefore, it is clear that studying LIRGs can shed new light on our understanding of the coeval evolution of host galaxy and black hole.

Mrk~938 (also known as NGC~34, NGC~17, and IRAS~F00085$-$1223, among others) is a local LIRG belonging to the {\it IRAS } Revised Bright Galaxy Sample \citep{Sanders2003} -- RA=00h11m06.65s, dec=--12d06m26.7s. It is located at a distance of 86.4\,Mpc using the recession velocity relative to the Local Group $cz_{\rm LG}$=5961\,km~s$^{-1}$ measured by \cite{Schweizer2007} and assuming a $\Lambda$CDM cosmology with ($\Omega_{\rm M}$,~$\Omega_{\Lambda}$)~=~(0.3,~0.7) and ${H}_{0}$~=~70~ ${\rm km}~{\rm s}^{-1}~{\rm Mpc}^{-1}$. The measured IR luminosity of this galaxy is $L_{\rm IR} = 3.4\times10^{11}\,{\rm L}_\odot$ using the {\it IRAS} fluxes reported by \citet{Sanders2003} and the adopted distance. Although this source has been classified in the literature as both Seyfert 2 and starburst (SB) galaxy from optical spectroscopy \citep[see][for a detailed discussion on this issue]{Schweizer2007}, its X-ray emission confirms the presence of an obscured AGN \citep[see][and also Section \ref{sec:AGN_X-ray}]{Guainazzi2005}. We note that in the recent work of \cite{Yuan2010} the nuclear activity of Mrk~938 has been definitively classified as Seyfert~2 based on optical spectroscopy.

\begin{figure}

\centering
\vspace{-0.8cm}
\hspace{-0.1cm}{\includegraphics[width=8.7cm,height=10.44cm]{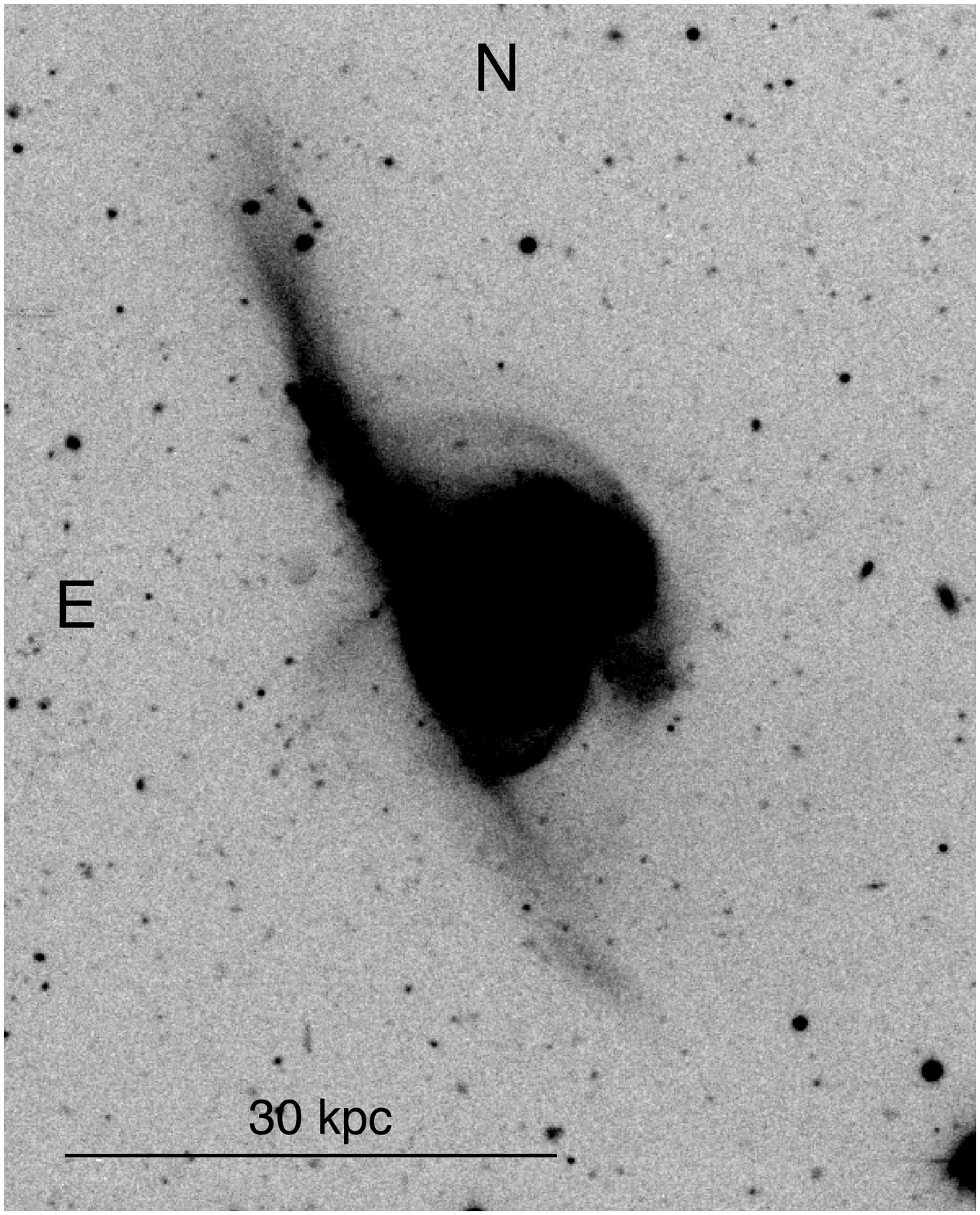}}

\vspace{-0.9cm}
\caption[Optical image]{Ground-based 2.5\arcmin $\times$ 3\arcmin\ B~band image of Mrk~938 provided by F.~Schweizer. Note the bright nucleus and spiral structure in addition to the two tidal tails emerging from the system in opposite directions.}
\label{fig:optical}

\end{figure}

The optical morphology of Mrk~938 is dominated by a bright red nucleus, a blue exponential disk, and a rich system of young massive star clusters. The faint optical surface brightness emission reveals a pair of tidal tails to the northeast and southwest of the system, with the former being the brightest and longest one (see Fig.~\ref{fig:optical}, courtesy of F.~Schweizer). All these signatures led \citet{Schweizer2007} to propose that Mrk~938 is the remnant of a gas-rich merger of two unequal mass galaxies. The nuclei of the two galaxies have probably coalesced, creating a concentrated and obscured starburst and an AGN, both driving a strong gaseous outflow. The starburst appears to be confined to a highly obscured central region of less than 1\,kpc in radius \citep{Schweizer2007}. This is indeed confirmed by ground-based high-angular resolution observations in the mid-IR. These revealed that the mid-IR emission in this system originates in the central $\sim 800\,$pc \citep{Miles1996}, but it actually dominates in a region of $\sim 400\,$pc (see discussion in Section \ref{sec:IRregion}). \citet{Fernandez2010} detected H~{\sc i} emission from both tidal tails and from nearby galaxies, suggesting that Mrk~938 is part of a gas-rich group of galaxies. They also found a hint of emission between Mrk~938 and NGC~35, the largest companion lying at a projected distance of 131\,kpc, and suggested that they might have recently interacted. The radio-continuum emission detected in the nuclear region is extended, indicating that it is mostly due to the highly concentrated starburst. 

Here we present for the first time a detailed analysis of novel far-IR ($70-500\,\mu$m) imaging observations of Mrk~938 obtained with the \herschel\ Infrared Observatory \citep{Pilbratt2010}. These have been combined with our own analysis of X-ray, UV, and IR data already presented in the literature. The main goal of this work is to understand the AGN/SB connection in Mrk~938, as well as the origin of the IR emission of this galaxy.  This is the second paper in a series presenting \herschel\ imaging observations of a sample of nearby Seyfert galaxies to investigate the co-existence of AGN and starburst activity, as well as the nature of the dusty torus. A study of the dust properties of NGC~3081 has recently been presented in \citet{Ramos2011b}.

This paper is structured as follows: Sect.~2 describes the data analysis procedure of the different IR and X-ray observations. In Sect.~3 we present evidence of AGN activity, including the decomposition of the mid-IR spectrum into AGN and SB contributions. The properties of the SB component have been derived in Sect.~4. We used the wealth of archival data to fit the SED of the galaxy and to derive the dust properties, presented in Sect.~5. Finally, our findings and conclusions are summarised in Sect.~6.

\section{Observations and data analysis}\label{sec:sec2}

\subsection{\herschel}
We performed  far-IR imaging observations of Mrk~938 with the  Photodetector Array Camera and Spectrometer \citep[PACS,][]{Poglitsch2010} and the Spectral and Photometric Imaging REceiver \citep[SPIRE,][]{Griffin2010} onboard the \herschel\ Space Observatory \citep{Pilbratt2010},
covering the spectral range $70-500$\,\micron\ in 6 bands. The data are part of the guaranteed time proposal ``Herschel imaging photometry of nearby Seyfert galaxies: testing the coexistence of AGN and starburst activity and the nature of the dusty torus'' (PI: M. S\' anchez-Portal).

The PACS observations were carried out using the ``mini-map'' mode, consisting of two concatenated scan line maps, at  70$^{\circ}$ and 110$^{\circ}$ (in array coordinates), at a speed of 20\arcsec/sec, each one with 10 lines of 3\arcmin~length and cross-scan step of 4\arcsec. This mode produces a highly homogeneous exposure map within the central 1\arcmin~area. The set of maps were duplicated to observe through both the 70\,\micron~(``blue'') and 100 \micron\ (``green'') filters. Therefore the galaxy was observed twice through the 160\,\micron~(``red'') filter. The PACS beams at 70, 100, and 160 \micron~are 5.6\arcsec, 6.8\arcsec, and 11.3\arcsec~FWHM, respectively. With the SPIRE photometer, the three available bands were observed simultaneously using the ``small map'' mode, with two 1\,$\times$\,1 nearly orthogonal scan lines, at a scan speed of 30\arcsec/sec. The scan line length is 11.3\arcmin~and the area for scientific use is around 5\arcmin$\times$5\arcmin. The FWHM beam sizes at 250, 350, and 500 \micron~are 18.1\arcsec, 25.2\arcsec, and 36.9\arcsec~respectively.

\begin{figure*}

\centering
\resizebox{0.9\hsize}{!}{\rotatebox[]{270}{\includegraphics{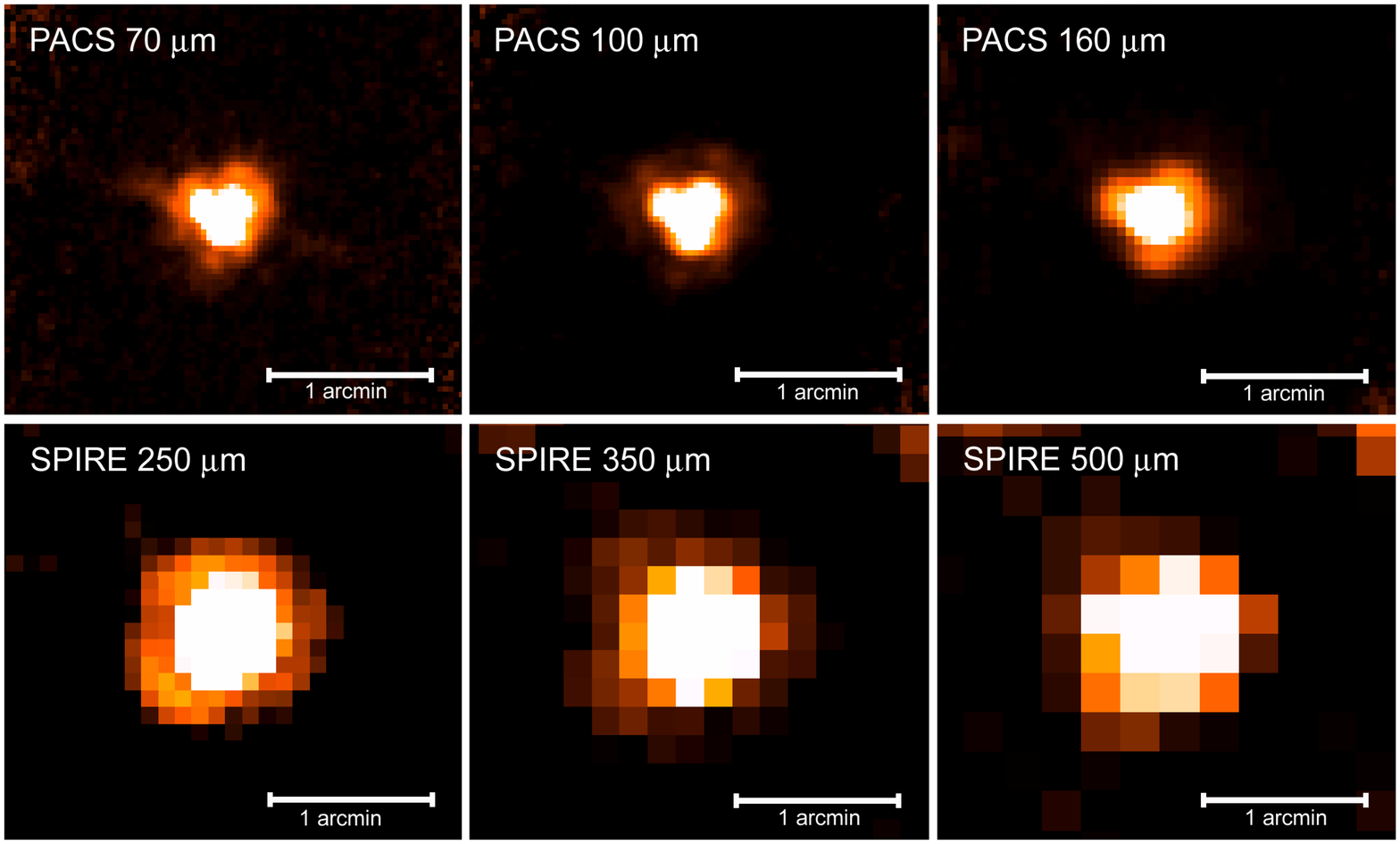}}}

\vspace{-3cm}
\caption{\herschel\ images of $3\arcmin \times 3\arcmin$ centred on Mrk~938. North/East is up/left respectively. Top: PACS 70, 100 and 160\,\micron; bottom: SPIRE 250, 350 and 500\,\micron; from left to right respectively. The PACS images show the shape of the PSF. In particular, a plume-like structure lying approximately in the NE-SW direction has been deemed as a PSF artifact and it should not be confused with the tidal tail observed in the IRAC images (Fig. \ref{fig:IRACmaps}).}

\label{fig:mosaic_image}
\end{figure*}

The data reduction was carried out with the \herschel\ Interactive Processing Environment (HIPE) v6.0.1951. For the PACS instrument, the extended source version of the standard {\tt PhotProject} reduction script was deemed as adequate, due to the small angular size of the galaxy. This procedure implements a high-pass filtering algorithm to remove the \textit{1/f} noise of the bolometer signals. The sources are masked in order to prevent the high-pass filter to remove flux from extended structures.  The FM v5 photometer response calibration files \citep{Muller2011} were used. Level~2 maps produced by the reduction script for the two scan directions were merged using the {\tt mosaic} task. For SPIRE, the standard small map script with the `na\"ive' scan mapper task was applied, using the calibration database v6.1. Colour corrections (for PACS, see Poglitsch et al.~2010; for SPIRE, please refer to the \textit{Observer's Manual}) are small for blackbodies at the expected temperatures \citep[e.g.][]{Perez2001} and have been neglected. More details on the processing of \herschel\ data are given in S\'anchez-Portal et al. (in preparation). The far-IR maps of Mrk~938 are shown in Fig.~\ref{fig:mosaic_image}.

\nocite{Poglitsch2010}

Aperture photometric data in the different bands were extracted using {\sc IRAF}\footnote{Image Reduction and Analysis Facility (IRAF) software is distributed by the National Optical Astronomy Observatories, which is operated by the Association of Universities for Research in Astronomy, Inc., under cooperative agreement with the National Science Foundation.} and the measurements for an aperture radius of 70\,\arcsec are shown in Table \ref{table:photometry}. The selected radius is a good compromise between including all IR emission and minimising the background contribution. Since the PACS encircled energy radius of FM v5 is normalised to an aperture of 60\,\arcsec, no aperture correction for the data was required. The morphology of Mrk~938, which will be further discussed in Sect.~\ref{sec:IRregion}, is compatible with an unresolved object. Using our far-IR and $L_{24\,\mu m}$ photometry and the parametrisation of \citet{Dale2002} we derived a total IR luminosity of $L_{\rm IR} = 3.3\times10^{11}\,{\rm L}_\odot$, in agreement with that of \citet{Sanders2003}.

\begin{table}
\centering
\begin{minipage}{140mm}
\caption{Integrated Infrared SED of Mrk~938.}
\label{table:photometry}
\begin{tabular}{@{}lcccc@{}}
\hline
Name &$\lambda$ (\micron)& $f_\nu$ (Jy)& error (Jy) & Reference\\

\hline
2MASS    & 1.2 &    0.051 & 0.001 & Jarret et al. (2003)\\
2MASS    & 1.6 &    0.067 & 0.002 & ``\\    
2MASS    & 2.2 &    0.062 & 0.002 & ``\\
\spitzer-IRAC     & 3.6 &    0.043 & 0.002 & Gallimore et al. (2010)\\
\spitzer-IRAC     & 4.5 &    0.036 & 0.001 & ``\\
\spitzer-IRAC     & 5.8 &    0.092 & 0.004 & ``\\
\spitzer-IRAC     & 8.0 &    0.59  &  0.01 & ``\\
\emph{IRAS}     & 12  &    0.35  & 0.032 & Sanders et al. (2003)\\	
\spitzer-IRS  & 15  &    0.435 & 0.012 & Deo et al. (2007)\\
\spitzer-MIPS     & 24  &    1.84  & 0.18  & This work\\
\emph{IRAS}     & 25  &    2.39  & 0.055 & Sanders et al. (2003)\\
\spitzer-IRS  & 30  &    4.36  & 0.05  & Deo et al. (2007) \\
\emph{IRAS}     & 60  &   17.05  & 0.045 & Sanders et al. (2003)\\
\herschel-PACS & 70  &   14.80  & 1.48  & This work\\
\emph{IRAS}     &100  &   16.86  & 0.135 & Sanders et al. (2003)\\	
\herschel-PACS &100  &   15.00  & 1.50  & This work\\
\emph{ISO}-ISOPHOT      &120  &    17.20 & 0.40  & Spinoglio et al. (2002)\\
\emph{ISO}-ISOPHOT      &150  &    10.60 & 0.50  & ``\\
\herschel-PACS &160  &     8.90 & 0.89  & This work\\
\emph{ISO}-ISOPHOT      &170  &     8.50 & 0.20  & Spinoglio et al. (2002)\\
\emph{ISO}-ISOPHOT 	 &180  &     5.40 & 0.10 & ``\\
\emph{ISO}-ISOPHOT 	 &200  &     3.00 & 0.10  & ``\\
\herschel-SPIRE &250  &     2.80 & 0.28 & This work\\
\herschel-SPIRE &350  &     0.90 & 0.09 & This work\\
\herschel-SPIRE &500  &     0.20 & 0.02 & This work\\
\emph{IRAM}-30m & 1300 & 0.0097 & 0.0018 & Albrecht et al. (2007)\\
\hline
\end{tabular}
\end{minipage}
\end{table}

\nocite{Deo2007}	
\nocite{Jarret2003}
\nocite{Spinoglio2002}
\nocite{Albrecht2007}

\subsection{\spitzer}
\subsubsection{Imaging Observations}
\begin{figure}
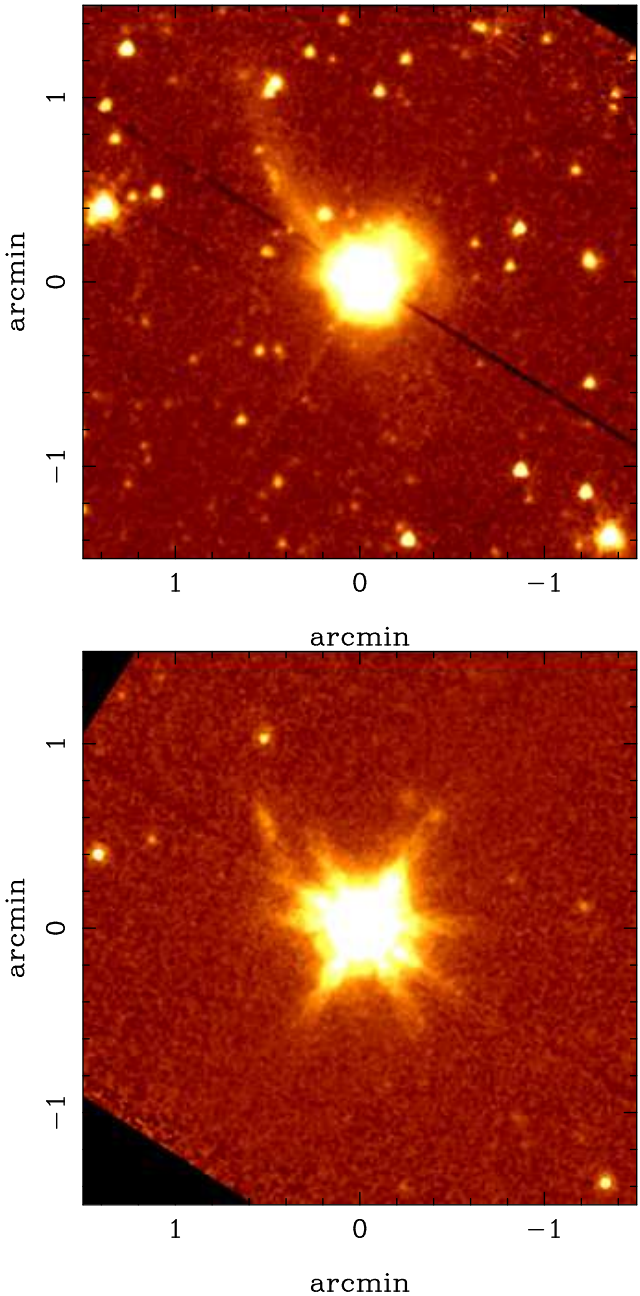


\centering
\resizebox{1.0\hsize}{!}{\rotatebox[]{0}{\includegraphics{IRAC36mu_map.ps}}}
\resizebox{1.0\hsize}{!}{\rotatebox[]{0}{\includegraphics{IRAC8mu_map.ps}}}

\caption[Spitzer images]{{\it Spitzer}/IRAC images at 3.6 (top) and 8\,\micron\ (bottom) of Mrk~938 covering a field of view of $3\arcmin \times 3\arcmin$. The orientation is north up, east to the left. Images are shown in a square root scale. Note that the northeast tidal tail and knots of star formation are clearly detected in the 3.6\,\micron\ image, which traces photospheric emission. In the 8\,\micron\ image, which probes dust continuun and PAH emission, one can see faint emission in that tail likely to be associated with the knots of star formation also seen in the optical (see Fig.~\ref{fig:optical}).}
\label{fig:IRACmaps}
\end{figure}

Mrk~938 was observed with the \textit{Spitzer} instruments IRAC \citep{Fazio2004} and MIPS \citep{Rieke2004}. We retrieved the basic calibrated data (BCD) from the \textit{Spitzer} archive (IRAC -- P3269, PI: J.F.~Gallimore; MIPS -- P3672, PI: J.M.~Mazzarella). The BCD processing includes corrections for the instrumental response (e.g. pixel response linearisation), flagging of cosmic rays and saturated pixels, dark and flat fielding corrections, and flux calibration based on standard stars (see the IRAC and MIPS instrument handbooks for details). We combined the BCD images into mosaics using the MOsaicker and Point source EXtractor (MOPEX) software provided by the \textit{Spitzer} Science Center (SSC) using the standard parameters. IRAC images of Mrk~938 at 3.6 and 8\,\micron\ are shown in Fig.~\ref{fig:IRACmaps}. The IRAC 8\,\micron\ image was affected by the banding effect (see IRAC Handbook). Similar to \citet{Gallimore2010} we corrected this artefact by fitting a polynomial to the affected rows. 

We computed integrated photometry of the source in the MIPS 24\,\micron\ image, and obtained from the literature photometric data points for the IRAC bands \citep{Gallimore2010}. The source photometry is presented in Table~\ref{table:photometry}.

\subsubsection{Spectroscopic Observations}
We retrieved \textit{Spitzer}\slash IRS \citep{Houck2004} spectroscopic observations of Mrk~938 from the {\it Spitzer} archive (P3269, PI: J.F.~Gallimore; P30291, PI: G.~Fazio). The observations were taken in low-resolution ($R\sim$60--120) with the Short-Low (SL) and Long-Low (LL) modules and in high-resolution ($R\sim600$) mode with the Short-High (SH) and Long-High (LH) modules. The low resolution data were observed in mapping mode while the high resolution data were obtained in staring mode.

For the staring data we started with the BCD data. Bad pixels were corrected using the {\tt IDL} package {\tt IRSCLEAN}\footnote{The {\tt IRSCLEAN} package is available at \\ \href{http://irsa.ipac.caltech.edu/data/SPITZER/docs/dataanalysistools/tools/irsclean/}{http://irsa.ipac.caltech.edu/data/SPITZER/docs/dataanalysistools/tools/irsclean/}}. Then we subtracted the sky emission and extracted the spectra using the standard programs included in the \textit{Spitzer} IRS Custom Extraction (SPICE) package provided by the SSC and the point source calibration. The resulting spectrum was already presented in \citet{Tommasin2010}. For the SL+LL mapping observations we constructed the data cubes using {\tt CUBISM} \citep{Smith2007}. We extracted the nuclear spectrum of Mrk~938 observed in SL+LL mapping mode using a 13\farcs4$\times$13\farcs4 aperture. Since the mapping mode data cubes are calibrated as extended sources  we applied a wavelength dependent aperture correction to obtain a point-like spectrum to represent the nuclear emission of this galaxy. For more details on the data analysis please see \citet{PereiraSantaella2010b,PereiraSantaella2010a,Alonso2011}.

We evaluated the strength of the 9.7\,\micron\ silicate feature, $S_{\rm Si}$, by calculating the ratio of the observed flux density to the continuum flux density at 9.7\,\micron\ using the method in \citet{PereiraSantaella2010a}. Applying the definition of \citet{Spoon2007} we get $S_{\rm Si}=-1.1$, where the minus sign indicates absorption. This value is considerably higher, in absolute terms, than the average of the LIRG sample in \citet{Alonso2011} of $\avg{S_{\rm Si}}=-0.6$. We also measured the equivalent width (EW) of the 6.2\,\micron\ PAH feature, and obtained EW(PAH~6.2\,\micron)=0.4$\pm 0.1$\,\micron. \citet{Spoon2007} provided a diagnostic diagram to give a general classification of IR galaxies based on $S_{\rm Si}$ and EW(PAH~6.2\,\micron). According to that, Mrk~938 is a moderately obscured composite nucleus.

\subsection{\xmm}\label{sec:xmm_obs}
Mrk~938 was observed by \xmm\ on 2002, December 22 (rev 556, obsid 0150480501, PI: R.~Maiolino) and then analysed with the \xmm\ Science Analysis Software \citep[SAS;][]{Gabriel2004}. All EPIC observations were taken in Full Frame mode and the medium filter in place. These data were already presented in \citet{Guainazzi2005}, but we reduced them with the latest software to take full advantage of the most recent calibration files. 

To extract the X-ray spectra, high background periods produced by intense soft proton fluxes were rejected by defining good time intervals within the complete observation window. After this process, the net exposure times were about 13, 16 and 16 ks for pn, MOS1 and MOS2 respectively. The light curves of the sources were inspected and no variability within the X-ray observations was observed. Source photons were extracted from a circular region of 30 arcsec radius centred on the object position. A circular source-free region on the same chip and radius of 60 arcsec was used to characterise the background. The target does not suffer from pile-up, so the pn spectrum was extracted with patterns 0 to 4, and MOS spectra from 0 to 12. Ancillary files and response matrices were generated by the {\tt arfgen} and {\tt rmfgen} SAS tasks respectively in order to convert the counts to physical units during the spectral analysis. All EPIC spectra were binned to oversample the instrumental resolution and to have no less than 20 counts in each background-subtracted spectral channel. 

\begin{figure}

\centering
\resizebox{1.0\hsize}{!}{\rotatebox[]{270}{\includegraphics{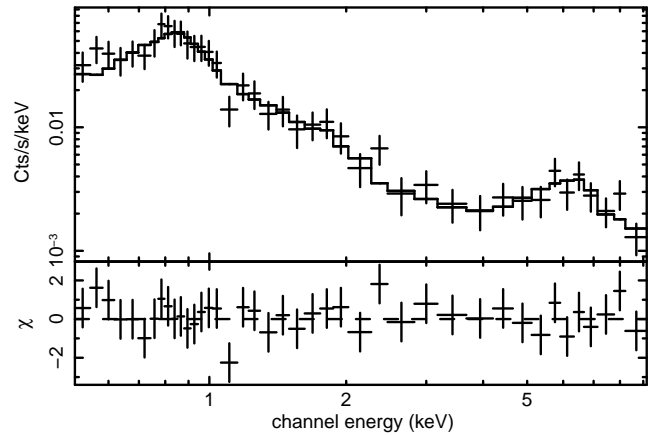}}}

\vspace{-1cm}
\caption[X-ray spectrum]{X-ray spectrum of the EPIC-pn data. The upper panel shows data and best-fit model (solid line). Residuals in terms of sigmas are plotted in the bottom panel.}
\label{fig:X-ray_spectrum}

\end{figure}

Spectral fitting was performed with {\tt XSPEC 12}, where all applied models include Galactic foreground absorption (${N}_{\rm H}$=$2.61\times10^{20}$ cm$^{-2}$) inferred from H~{\sc i} observations \citep{Dickey1990}. Given that our source has been found to be highly obscured, we started fitting the data with a model consisting of a powerlaw plus an absorbed powerlaw, both with the same photon index. We got a reasonable fit at high energies, but a prominent soft excess with respect to the powerlaw appeared below $\sim$1\,keV. Thus we added a thermal component, {\tt MEKAL} in {\tt XSPEC}. We obtained a temperature $kT$=0.67\,keV typical of a starburst emission, and a high intrinsic absorption of $N_{\rm H}^{int}$=75$^{+31}_{-21}\times 10^{22}$ cm$^{-2}$, consistent with high obscuration. A considerably flat powerlaw was obtained, \slope=1.43. A summary of the parameters of this model, which is considered our best-fit ($\chi^2$/dof=50/64), is presented in Table \ref{table:Xray_fitting}. Our derived X-ray luminosity is similar to that in \citet{Guainazzi2005}. Fig. \ref{fig:X-ray_spectrum} shows the X-ray data (only EPIC-pn data is shown for simplicity) and model-to-data residuals. We also tried adding a reflection model \citep[{\tt PEXRAV}, ][]{Magdziarz1995}, for possible X-ray scattering from an optically thick surface, but did not get any statistical improvement in the fit.

\begin{table}

\caption{Parameters derived from the fitting of the best-fit model, zwabs$\times$(mekal+zpo+zwabs$\times$zpo) in {\tt XSPEC} nomenclature, to the \xmm\ data. Galactic foreground absorption has been used for the first model component. Luminosities have been corrected for absorption. Errors are 90\% confidence.}

\label{table:Xray_fitting}

\begin{tabular}{l c c c c c c }
\hline 

\textbf{$kT$} & 
\textbf{$N_{\rm H}^{int}$} &
\textbf{\slope} &
\textbf{$L_{\rm 0.5-2\,keV}$} &
\textbf{$L_{\rm 2-10\,keV}$} \\

(keV) &  (10$^{22}$ cm$^{-2}$) & & ($10^{40}$\,\lumunit) & ($10^{42}$\,\lumunit) \\ 

\hline

0.67$^{+0.08}_{-0.07}$	& 75$^{+31}_{-21}$	& 1.43$^{+0.24}_{-0.26}$	& 5.4$^{+0.9}_{-0.8}$	&	1.4$^{+0.3}_{-0.2}$	\\

\hline
\end{tabular}
\end{table}

\subsection{\emph{Swift}}

The position of Mrk~938 has been observed with the Burst Alert Telescope \citep[BAT][]{Barthelmy2005}, but the source is not detected even in the 70th month survey. We have used the standard BAT tool {\tt batcelldetect}\footnote{\href{http://heasarc.gsfc.nasa.gov/ftools/caldb/help/batcelldetect.html}{http://heasarc.gsfc.nasa.gov/ftools/caldb/help/batcelldetect.html}} to analyse the observations and derive the rate (normalised by the number of working detectors) at the source position in the eight standard channels from corresponding background subtracted images. In the fourth channel, 35--50\,keV, the rate becomes negative, therefore we have combined the first three channels to get a rate in the 14$-$35\,keV energy range. Using {\tt WebPIMMS}\footnote{\href{http://heasarc.nasa.gov/Tools/w3pimms.html}{http://heasarc.nasa.gov/Tools/w3pimms.html}} for the count rate to flux conversion we get a flux upper limit of 2.5$\times 10^{-12}$\,\fluxunit~(equivalent to a luminosity of 2.2$\times 10^{42}$\,\lumunit). If we extrapolate our best-fit model derived for the XMM-Newton data, we get $F_{\rm 14-35\,keV}$ =1.6$^{+0.8}_{-0.4}\times 10^{-12}$\,\fluxunit ($L_{\rm 14-35\,keV}$ =1.4$^{+0.9}_{-0.5}$~$\times10^{42}$\,\lumunit), that is compatible with a BAT non-detection. We cannot further constrain the X-ray model because the extrapolated fluxes for all reasonable models lie below the BAT upper limit.

\section{AGN Activity}

In the local universe, a significant fraction of LIRGs hosts an AGN, and this fraction increases with the IR luminosity
of the system \citep{Veilleux1995}. Using optical emission-line ratios and a new scheme to classify galaxies, \citet{Yuan2010} recently reported that approximately half of local LIRGs are classified as Seyfert, LINER or composite (SB/AGN). The AGN detection rate in LIRGs 
can be as high as $\sim 60\%$ when combining optical and IR indicators \citep[i.e. emission line ratio diagnostics and spectral decomposition,][]{Alonso2011}, and thus similar to that of local ultraluminous infrared galaxies \citep[ULIRGs, $L_{\rm IR}=10^{12}-10^{13}\,L_\odot$][]{Nardini2010}. However, in clear contrast with ULIRGs, in local LIRGs the AGN bolometric contribution to the IR luminosity of these systems is very small, with an average of $\avg{L_{\rm bol}({\rm AGN})/L_{\rm IR}}$=0.05$^{+0.07}_{-0.03}$ \citep{Alonso2011}. Therefore the high IR luminosities in most LIRGs are produced in intense dusty starbursts. \citet{Petric2011} estimated that AGN contribute $\sim$12\% to the IR luminosity, which is compatible with the upper end found in \citet{Alonso2011}.

\subsection{X-ray evidence}\label{sec:AGN_X-ray}
 
The analysis of the X-ray emission of Mrk~938 shows evidence of the presence of an obscured AGN. The active nucleus dominates the hard X-ray contribution as can be seen from the highly absorbed powerlaw component used to fit the spectrum. The measured absorption-corrected hard X-ray luminosity of $1.4^{+0.3}_{-0.2}\times 10^{42}$\,\lumunit~is too high to be originated in a pure starburst (see Sect \ref{sec:SB_X-ray}). The high intrinsic absorbing column returned by the spectral fitting supports the conclusion of the highly obscured nucleus.

A powerful diagnostic to estimate the nuclear obscuration is the 2$-$10 keV to [O\,{\sc iii}] flux ratio, $T$ \citep[e.g.][]{Maiolino1998, Bassani1999, Guainazzi2005}. According to previous studies the threshold lies in \emph{T}=0.1, where objects with $T\leq$0.1 are Compton-thick while those with $T\geq$1 are unobscured or Compton-thin. Using $F_{[\rm O\,\textsc{iii}]}$=1.1$\times 10^{-14}$\,\fluxunit\,as presented in \citet{Polletta1996} we obtain T=0.01 for our source, therefore Mrk~938 would appear as Compton-thick. However, this galaxy has a very strong starburst contribution in the optical \citep{Gonsalves1999}, therefore the AGN is not responsible for a considerable fraction of the [O\,{\sc iii}] flux and the $T$ ratio as a measure of obscuration is uncertain. Another indicator of the nuclear obscuration is the emission fluorescence  Fe K$_\alpha$ line. Objects are considered as Compton thick when EW(Fe K$_\alpha$)$>$1000\,eV \citep{Matt1996}. In the case of Mrk~938, due to low statistics at the hardest energies we can only constrain the presence of the Fe K$_\alpha$ line by extracting an upper limit. Using Cash statistics, \citet{Guainazzi2005} fitted the unbinned spectrum in the 5.25--7.25\,keV range. They found an equivalent width upper limit of 322\,eV when fitting a Gaussian with energy fixed at 6.4\,keV. Therefore, the width of the Fe K$_\alpha$ line places the source in the Compton-thin region.

\subsection{Mid-infrared evidence}\label{sec:MIR_evidence}

In terms of continuum emission, the mid-IR spectral range contains a wealth of information to disentangle the AGN from the SB contribution in galaxies. This comes from both broad features and fine structure emission lines. In particular, high metallicity starburst galaxies show very similar mid-IR spectra \citep[see e.g.,][]{Brandt2006, Bernard2009}, which are rather different from the continuum and line emission observed in pure AGN. Because of this, spectral decomposition of {\it Spitzer} IRS spectra is a powerful tool to estimate the fractional AGN and SB contributions at different mid-IR wavelengths in both local \citep[see e.g.,][]{Nardini2008, Nardini2010, Alonso2011} and high-$z$ \citep[e.g.][]{Menendez2009} galaxies.

In the case of local LIRGs \citet{Alonso2011} demonstrated that the mid-IR spectral decomposition can identify even subtle AGN emission that may otherwise be under the detection threshold with other mid-IR diagnostics. Here we used a spectral decomposition method very similar to theirs to estimate the AGN contribution with respect to the total mid-IR emission of Mrk~938. For the SB component this method uses the average starburst spectrum of \citet{Brandt2006}, and the templates of purely star forming (U)LIRGs of \citet{Rieke2009} covering the IR luminosity range 10.5\,$\leq \log (L_{\rm IR}/{\rm L}_\odot)\leq 12$. The AGN continuum emission is represented  by that produced by an interpolated version of the {\tt CLUMPY} torus models of \citet{Nenkova2008a, Nenkova2008b}. We restricted the large database of the {\tt CLUMPY} torus models to only use two AGN templates. These were derived by \citet{Ramos2011} as the best fitted models to the average SEDs of nearby type 1 and type 2 Seyferts using a Bayesian code \citep{Asensio2009}. We then looked for the best combination of SB + AGN templates to fit the IRS spectrum that minimised $\chi^2$. The best fit to the IRS spectrum of Mrk~938 was achieved with the type 2 AGN template, as expected, plus the $10^{\rm 12}\,{\rm L}_\odot$ starburst template. This SB template is required to fit the relatively deep 9.7\,\micron\ silicate feature observed in this galaxy. We note that the IR luminosity of the best fit template is higher than that of our galaxy. This might indicate that the IR emission comes from a relatively compact region (Sect.~\ref{sec:IRregion}), as is the case of many LIRGs.

\begin{figure}

\hspace*{-0.5cm}
\resizebox{1.05\hsize}{!}{\rotatebox[]{0}{\includegraphics{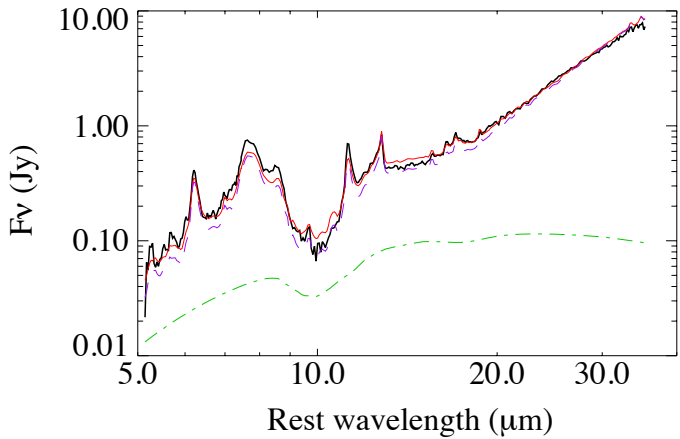}}}

\vspace{0cm}
\caption[MIR fir]{AGN+SB best fit of the observed SL+LL IRS spectrum of Mrk~938 (black line). The dashed-dotted green line is the fitted scaled Seyfert 2 torus model template (see text for details), the purple dashed line is the fitted scaled SB template, and the solid red line is the sum of the fitted SB and AGN components.}

\label{fig:mir_fit}
\end{figure}

In Fig.~\ref{fig:mir_fit} we show the IRS spectrum of Mrk~938 plus the AGN+SB best fit model, in addition to the scaled model components. As can be seen from the figure, the AGN contribution to the mid-IR emission is small but detectable. The AGN fractional contributions to the continuum emission within the {\it Spitzer}/IRS slit are listed in Table~\ref{table:AGN_contribution}. Although our statistical uncertainties on the derived parameters are small ($\pm$5\%) the errors associated to the templates can be larger, especially in the SB component. Our results are in good agreement with those of \citet{Deo2009}. Clearly, as found in local LIRGs and many local Seyfert galaxies, the AGN contribution at wavelengths longer than 40\,\micron\ is greatly reduced \citep[see][]{Alonso2011,Mullaney2011}. \citet{Vega2008} obtained a contribution of 17\% of the AGN to the 3--30\,\micron\,luminosity, which is at the upper end of the AGN contribution that we find at 20\micron\ including uncertainties. \citet{Tommasin2010} find an AGN contribution to the 19\micron\ flux of Mrk 938 of 50\,($\pm$13)\%, which is considerably higher than our estimation. This can be explained because they used upper limits on the undetected line widths for their estimation.

\begin{table}

\caption{AGN fractional contributions within the IRS slit obtained from the AGN+SB decomposition.}

\label{table:AGN_contribution}

\begin{tabular}{c c c c}
\hline 

$C^{\rm IRS}_{6\,\micron}$[AGN] & 
$C^{\rm IRS}_{20\,\micron}$[AGN] &
$C^{\rm IRS}_{24\,\micron}$[AGN] &
$C^{\rm IRS}_{30\,\micron}$[AGN] \\

\hline

0.21	& 0.10	& 0.05	& 0.02		\\

\hline
\end{tabular}
\end{table}

Additionally, we can use mid-IR high excitation emission lines to investigate the AGN emission, in particular, the [Ne~{\sc v}] lines at 14.32\,\micron\ and 24.32\,\micron\ and the [O\,{\sc iv}]~$\lambda$25.89\,\micron\ line \citep{Genzel1998}. \citet{Tommasin2010} did not detect any of these emission lines in Mrk~938 \citep[see also][]{Gallimore2010, PereiraSantaella2010b}. When these high excitation emission lines are ratioed with a line mostly produced by star formation (e.g., the [Ne\,{\sc ii}]~$\lambda$12.81\,\micron), it is possible to obtain an approximate contribution -if any- of the AGN to the mid-IR emission of a galaxy. The upper limit to the [O\,{\sc iv}]~$\lambda$25.89\,\micron/[Ne\,{\sc ii}]~$\lambda$12.81\,\micron\ of $\sim 0.01$ \citep{Tommasin2010, PereiraSantaella2010b} indicates that the mid-IR emission of this galaxy is  dominated by star formation activity \citep[see e.g.,][]{Alonso2011,Petric2011}. This is in good agreement with the low AGN fractional contribution to the mid-IR emission inferred from the spectral decomposition.

Since the IRS slit encloses most of the 24\,\micron\ emission (from the comparison with the MIPS photometry, see Table~\ref{table:photometry}),  the derived AGN fraction contribution to the 24\,\micron\ emission is 0.05. \citet{Alonso2011} also found that in local LIRGs this fractional AGN contribution is related to the AGN bolometric contribution to the IR luminosity of the system. Using their relation we estimated that in Mrk~938 $L_{\rm bol}({\rm AGN})/L_{\rm IR} = 0.02^{+0.02}_{-0.01}$. This is in good agreement with the AGN bolometric contribution estimated by \cite{Vega2008} of 4\% from their modelling of the full SED of this galaxy. We derived then an AGN bolometric luminosity of 2.6$^{+2.6}_{-1.3}$ $\times 10^{43}$\,\lumunit. Alternatively, the AGN bolometric luminosity can be estimated from the hard X-ray $2-10\,$keV luminosity. Using the bolometric correction of \citet{Marconi2004} and our derived X-ray luminosity from the model fit $L_{\rm 2-10\,keV}=1.4\times 10^{42}$\,\lumunit (Table~\ref{table:Xray_fitting}), we get a bolometric luminosity of $L_{\rm bol}({\rm AGN})$=1.6$^{+1.1}_{-0.6}$ $\times 10^{43}$\,\lumunit, compatible with our previous estimation derived from the IRS spectral decomposition.

\section{Starburst Activity}
It is now becoming clear that the size of star-forming regions and the resulting IR surface brightness (also referred as compactness)
of IR bright galaxies may help us understanding how star formation is produced in galaxies. That implies that we may be able to differentiate between extended steady star formation in disks and compact star formation activity driven by mergers. These two processes produce different overall IR spectral energy  distributions (SEDs) of LIRGs and ULIRGs both locally and at high-redshift \citep[see e.g.,][and references therein]{Rujopakarn2011, Rujopakarn2011b}, where there is no clear dependence of the shape of SED with $L_{\rm IR}$.

Similar to the fundamental planes of spiral and elliptical galaxies, there are three parameters that define a fundamental plane of IR-selected galaxies. \citep{Chanial2007}. These are: the IR luminosity, the dust temperature and the size of the IR-emitting regions. According to these authors, changes in the IR compactness describe a smooth sequence going from quiescent to starburst galaxies, where there is a simultaneous increase of gas surface density, effective dust temperature, and SFR surface density. In the following sections we explore in detail some of these properties in Mrk~938.

\subsection{IR Emitting Region}\label{sec:IRregion}

The detailed study of the optical properties of Mrk~938 (morphology, star cluster ages, and dynamics) led \cite{Schweizer2007} to claim that this object is the result of a merger of two gas-rich galaxies of unequal masses (1/3$\lesssim m/M \lesssim$2/3). At present, the current starburst activity of Mrk~938 appears to be confined to a highly obscured central region of less than 1\,kpc in size.

Fig.~\ref{fig:IRACmaps} presents the {\it Spitzer}/IRAC images at 3.6 and 8\,\micron. The 3.6\,\micron\ image, which probes the photospheric emission, shows the main body of the galaxy plus some extended low surface brightness emission. The northeast tidal tail of this merger system is clearly detected in the form of diffuse emission and some knots. There also exists a southwest tail with a much fainter optical surface brightness that is hardly seen in the 3.6\,\micron\ image. The knots in the northeast tail, also detected in the IRAC 8\,\micron\  image, appear to be coincident with the blue knots lying at a projected distance of $\sim 50\arcsec$ ($\sim 21\,$kpc) from the nucleus \citep{Schweizer2007}. Knots in both tidal tails are associated with H~{\sc ii} regions (A.~P\'{e}rez-Garc\'{\i}a, private communication).

The IRAC 8\,\micron\ emission, which probes dust continuum emission and PAH emission, is dominated by the presence of a bright central region. This region appears slightly resolved with a size (FWHM) of $2.3\arcsec$ ($\sim 950\,$pc). The size of the 8\,\micron\ emitting region inferred from the IRAC data is in good agreement with the distribution of the mid-IR emission detected from ground-based high angular resolution imaging as measured by \citet{Miles1996}. These authors also showed that there are two mid-IR sources in the central regions of Mrk~938 separated 1.2\,\arcsec, although most of the mid-IR emission arises from a resolved source with an angular size of $1\arcsec$ (FWHM). This confirms that the starburst region in this galaxy is only relatively compact ($\sim 0.5-1\,$kpc, FWHM), as is the case in a large fraction of local LIRGs \citep{Alonso2006, Rujopakarn2011}. Also, the presence of extended mid-IR emission in the central regions of Mrk~938 is in agreement with the result that the AGN contribution to the mid-IR emission is low, as demonstrated in Sect.~\ref{sec:MIR_evidence}.  We used PACS 70\,\micron\ and MIPS 24\,\micron\ images, that have comparable angular resolution, and probed the current star formation. We can set an upper limit to the starburst region of $\le 5\arcsec \sim 2\,$kpc (FWHM), which is in good agreement with the measured size at shorter wavelengths.

We analysed the {\it Herschel} brightness profiles of Mrk~938 and compared them with those of point-like sources. We found no evidence of extended emission in any of the PACS or SPIRE bands. As a sanity check, we performed aperture photometry on the PACS 70\,\micron~image using small radii, in this case 6 and 14\,\arcsec. After applying the corresponding aperture corrections we obtained values in agreement with the integrated flux at 70\,\micron, therefore confirming that we are not detecting extended emission.

\subsection{Star Formation Rate}\label{sec:SFR}

Based on a variety of morphological, photometric and spectroscopic arguments, \cite{Schweizer2007} concluded that the star formation activity of Mrk~938 was higher and more spread out in the past. As a result of the merger process a galaxy-wide starburst started about 600 million years ago. It is this post-starburst population that appears to dominate the emission from the blue disk present in this system. \cite{Riffel2008} used near-IR spectroscopy to study the stellar populations in the central region of this galaxy. They found that the near-IR emission of this galaxy is dominated by an intermediate-age stellar population of $\sim 1\,$Gyr, whereas the {\it young} ($\le 30\,$Myr) stellar population contributes to less than 25\% of this emission. 

In the previous section we saw that most of the IR emission of Mrk~938 appears to be coming from a relatively compact region 0.5$-$2\,kpc in size. In very dusty galaxies, as is the case of Mrk~938, this IR emission is probing the {\it current} star formation rate (SFR). This is because the UV light emitted by the young stars  is absorbed by the dust and re-emitted in the IR. Therefore, some IR monochromatic luminosity or, alternatively, the total IR luminosity can be used to estimate the SFR \citep{Kennicutt2009}. 
The 24\,\micron\ monochromatic luminosity has been found to be a good indicator of the current SFR of dusty galaxies \citep[e.g.][]{Alonso2006, Calzetti2007, Rieke2009}. It has the advantage that it traces dust heated by the most massive stars, hence directly probing the \emph{current} star formation  activity. We used the \cite{Rieke2009} calibration

\begin{eqnarray}
{\rm SFR_{IR}} \left(M_\odot\,{\rm yr}^{-1}\right)&=&7.8 \times 10^{-10} L_{24\mu \rm m} \left(L_\odot\right) \notag \\
& & \times \{7.76 \times 10^{-11} L_{24\mu \rm m}\left(L_\odot\right)\}^{0.048}
\end{eqnarray}

\noindent
\noindent
which assumes a Kroupa IMF\footnote{Using a Salpeter IMF would result in a SFR approximately 1.4 times higher than that from the Kroupa IMF \citep[see e.g.][]{Kennicutt2009}.}. Using $L_{24\,\mu \rm m}=5.3\times10^{10}$\,\Lsun\,(calculated from the 24\,\micron\ flux in Table~\ref{table:photometry}), we estimated a SFR$_{\rm IR}$ of 42\,\msun~yr$^{-1}$ to an accuracy within 0.2\,dex (after removing the AGN contribution at this wavelength).  This value is actually the {\it obscured} SFR. 
This value lies between the values estimated by \citet{Vega2008} of 21 and 151\,\msun~yr$^{-1}$ based on SED fitting for star formation in the last 10\,Myr and averaged over the entire burst respectively. Also, our derived SFR is similar to the 64\,\msun~yr$^{-1}$ obtained from 1.4\,GHz observations \citep{Fernandez2010}. As a sanity check we can test the above hypothesis that most of the on-going SFR in LIRGs is taking place in obscured environments.

A source of uncertainty in the SFR calculation using IR indicators arises if a significant fraction of the stellar luminosity is not absorbed and then re-emitted in the IR, but escapes directly in the UV instead. We derived the SFR$_{\rm UV}$ using the far-UV flux for this galaxy. This was found by correlation of the source position with the GALEX archive. Taking into account a Galactic colour excess $E(B-V)$=0.027 (quantity reported in NED\footnote{The NASA/IPAC Extragalactic Database (NED) is operated by the Jet Propulsion Laboratory, California Institute of Technology, under contract with the National Aeronautics and Space Administration}) and applying the extinction curve of \citet{Fitzpatrick1999} we derived the UV corrected flux ($F_{\rm UV}^{c}$=0.62\,mJy). Following \citet{Kennicutt1998}, assuming a flat continuum and scaling to a Kroupa IMF we have

\begin{eqnarray}
{\rm SFR_{UV}} \left(M_\odot\,{\rm yr}^{-1}\right) = 9.2 \times 10^{-29} L_{\nu}~\left({\rm erg~s}^{-1}\right)
\end{eqnarray}

\noindent
where $L_{\nu}$ is the UV luminosity. For Mrk~938 we find that the ${\rm SFR_{UV}}$ represents $\sim$3\% of the ${\rm SFR_{IR}}$, as also found in \citet{Howell2010}. This value is in agreement with the median value found for a sample of LIRGs in \citet{Buat2007} and \citet{Pereira2011}.

Similarly, we can estimate the {\it unobscured} SFR from the observed (not corrected for extinction) integrated H$\alpha$ luminosity ($L_{\rm H\alpha}=10^{42}$\,\lumunit) reported by \cite{Moustakas2006}. We used the SFR recipe of \cite{Kennicutt2009} that accounts for the {\it unobscured} and {\it obscured} SFR in terms of the H$\alpha$ and 24\,\micron\ luminosities, respectively. We inferred that most ($\sim 90-95\%$) of the current star formation in Mrk~938 is {\it obscured}, and thus likely to be taking place in the compact starburst region revealed by the mid-IR observations described above. This is in agreement with the small value found for the ${\rm SFR_{UV}}$ calculated above.

\subsection{X-rays from star formation}\label{sec:SB_X-ray}

For a starburst, the emission of radiation in the X-ray domain can be attributed to a combination of components like supernova remnants, hot plasmas associated to star-forming regions and High Mass X-ray Binaries (HMXB) within the galaxy among others \citep{Fabbiano1989}. Several investigations have revealed correlation between the X-ray emission and total SFR in starbursts \citep[e.g.][]{Ranalli2003,Pereira2011}. The linear relation of \citet{Pereira2011} reads

\begin{eqnarray}\label{eq:eq3}
{\rm SFR_{IR+UV}} \left(M_\odot\,{\rm yr}^{-1}\right) = {constant} \times 10^{-40} L_{\rm X}~{\rm erg~s}^{-1}
\end{eqnarray}

\noindent
with a 0.25 dex scatter, where $constant$=3.4$(\pm 0.3)$ and 3.9 $(\pm 0.4)$ for the soft 0.5--2\,keV and hard 2--10\,keV X-rays respectively. Applying this equation with the SFRs derived in Sect.~\ref{sec:SFR}, we estimate the following luminosities due to star formation activity in Mrk~938: $L_{\rm 0.5-2\,keV}=1.2 (\pm 0.6) \times 10^{41}$ \,\lumunit and $L_{\rm 2-10\,keV}=1.0 (\pm 0.6) \times 10^{41}$ \,\lumunit. The errors take into account the uncertainties in the SFR value and the uncertainties in Eq.~\ref{eq:eq3}.

We have compared the X-ray emission expected from star formation activity as calculated above with that of the fitted \xmm\ data.
Using the luminosities measured in Sect.~\ref{sec:xmm_obs} for the best-fit model to the X-ray data we get a consistent value for the soft component (0.5$^{+0.2}_{-0.2} \times 10^{41}$\,\lumunit). This is in agreement with the fact that most, if not all, of the X-ray emission at soft energies is produced by the starburst, as is also the case in a large fraction of local LIRGs \citep{Pereira2011}. Also, the model derived from the fit of the X-ray data is consistent with star formation. However, in the 2--10\,keV band, the luminosity of the X-ray fitting is an order of magnitude higher than that expected from the SFR. This was \emph{a priori} anticipated because the AGN is the major contributor to the harder energies.

\section{Spectral Energy Distribution and Dust Properties}\label{sec:SED}

Several attempts have been made in the past to model the IR and
submillimeter spectral energy distributions (SED) of LIRGs and ULIRGs
using a modified blackbody with one or several dust temperature  or
full radiative transfer modelling \citep[see e.g.][]{Klaas2001,
  Dunne2001, Vega2008, Clements2010}. Here we use for the first time
{\it Herschel} photometry  and data from the literature to obtain
better estimates of the dust properties of
Mrk~938. Table~\ref{table:photometry} presents our {\it Herschel}
photometry as well as {\it IRAS}, {\it Spitzer} and {\it ISO} data
compiled from the literature to construct the far-IR SED. The errors
quoted for the {\it Herschel} flux densities take into account the
photometric calibration uncertainties of $\sim 10\%$ in all the bands.
The remaining errors listed in
Table~\ref{table:photometry} are those given in the corresponding
references.

\begin{figure*}
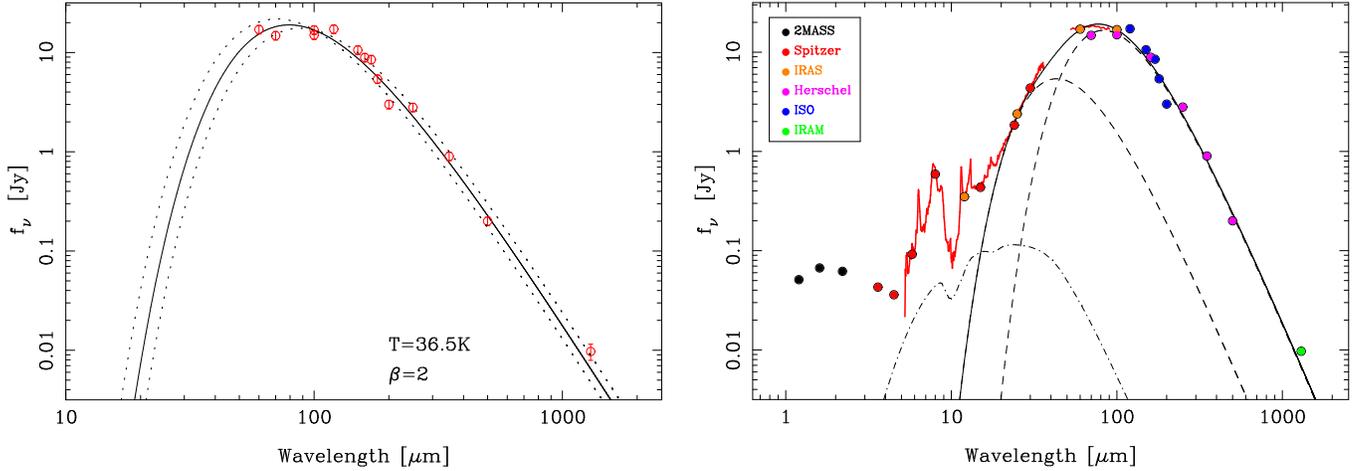


\centering

\begin{tabular}{cc}
\put(-250,50){\includegraphics[height=6.2cm]{Fig_SEDFIR_Mrk938.ps}}
\put(10,14){\rotatebox[]{270}{\includegraphics[width=6.2cm]{Mrk938_fullSED.ps}}}
\end{tabular}

\vspace{-1.5cm}
\caption[SED]{SEDs of Mrk~938. The left panel shows the far-IR SED
  and the best-fit (solid line) with a modified blackbody with
  fixed $\beta=2$ and 
  a best-fit temperature of $T_{\rm dust}=36.5$\,K. 
We also plotted acceptable fits (see text) as dotted lines,
  correspoding to $T_{\rm dust}=33\,$K and $T_{\rm dust}=41\,$K.
The right panel
  shows the full IR SED including all the photometric points in
  Table~\ref{table:photometry} (the colour code for the different
  instruments appears in the figure inset), as well as our extracted
  IRS spectrum and the MIPS-SED data of \cite{Gallimore2010} as red
  solid lines. We have also plotted our modelling to the mid and far-IR data
  using two modified blackbodies (solid black line), with fixed
  $\beta=2$ and best-fit temperatures of $T_{\rm dust}^{\rm w}=67$\,K
  and $T_{\rm dust}^{\rm c}=35$\,K for the warm and cold components,
  respectively (dashed lines). The grey dotted-dashed line is the
  fitted type 2 AGN template derived in
  Section~\ref{sec:MIR_evidence}.}. 
        \label{fig:SED}
    \end{figure*}

We parametrised the SED of Mrk~938 using the standard modified blackbody, which depends on the dust temperature ($T_{\rm dust}$) and the dust emissivity ($\beta$). In the case of optically thin emission in the IR, this function can be approximated as:

\begin{equation}
f_{\nu} \propto \frac{\nu^{3+\beta}}{{\rm e}^{h\nu/KT_{\rm dust}}-1}
\end{equation}

We fixed $\beta=2$ as it has been found to be a reasonable value 
  for IR bright galaxies \citep{Dunne2001}. Then we used a $\chi^2$
minimization 
method to fit the dust temperature. The SED fit was done using all the
flux densities for $\lambda \ge 60$\,\micron\ given in 
Table~\ref{table:photometry}, with the normalization at 100\,\micron. For the {\it IRAS} and {\it ISO} flux densities we used a 10\% uncertainty.

We obtained a best-fit dust temperature of $T_{\rm dust}= 36.5\,$K. As
can be seen from Fig.~\ref{fig:SED} (left panel), using a single 
temperature produces a good fit to the far-IR SED of Mrk~938.   
Acceptable fits ($\chi^2< 1.5\times \chi^2_{\rm min}$) were also
obtained in the range of temperatures  $T_{\rm dust}=36^{+4}_{-3}\,$K
(the fits for the limitting temperatures are shown in 
Fig.~\ref{fig:SED} as dotted lines).
The best-fit  temperature is similar to those found by
\cite{Klaas2001} for local (U)LIRGs using similar assumptions. We also
tried a fit using $\beta=1.5$, which has been found  to fit better the
SEDs of LIRGs and ULIRGs \citep{Dunne2001, Clements2010}. The fit
provided a higher dust temperature, as expected. However, this fit was
considerably worse, having a value of $\chi^2$ more than three times higher than
that obtained with $\beta=2$.

We derived the dust mass ($M_{\rm dust}$) using the following equation \citep[adapted from][]{Hildebrand1983}:

\begin{equation}
M_{\rm dust} = \frac{D_{\rm L}^2 f_\nu}{\kappa_\nu B_\nu(T_{\rm dust})}
\end{equation}

\noindent
where $f_\nu$ is the observed flux density, {\rm $D_{\rm L}$ is the
  luminosity} distance, $B_\nu(T_{\rm dust})$ is the blackbody
emission for the best-fit 
dust temperature, and $\kappa_\nu$ is the absorption coefficient. We
evaluated this expression at 250\,\micron\ using an absorption
coefficient of $\kappa_{250\mu{\rm m}}=4.99\,{\rm cm}^2\,{\rm g}^{-1}$
as interpolated from the dust model of \cite{Li2001}.  We obtained
$M_{\rm dust}=3\times 10^7\,{\rm M}_\odot$, which is consistent with
the values derived for local ULIRGs and other IR-bright galaxies
\cite[see][and references therein]{Klaas2001, Clements2010}.  For
  the range of acceptable dust temperatures, the inferred uncertainty
  in the calculated dust mass is $\sim 20\%$. \citet{Vega2008} 
 estimated an $M_{\rm dust}$ of $8.5\times 10^7\,{\rm M}_\odot$. However, they used the dust-to-mass ratio as conversion factor between the gas and the dust mass, using a fixed value of 100 for all sources in their sample. This can introduce a large uncertainty in the dust mass. This is most probably the reason for the difference between our value and that in \citet{Vega2008}.

We have also fitted the SED of LIRGs using a two-temperature model,
similar to those used in other works
\citep[e.g.][]{Dunne2001,Clements2010}, where we have taken into
account all flux density values for $\lambda \ge 24$\,\micron\ to
obtain a better constraint of the properties of the warm
dust. Using two modified blackbody functions 
and $\beta$=2 we obtained $T_{\rm dust}^{\rm w}=67$\,K and $T_{\rm
  dust}^{\rm c}=35$\,K for the warm and cold components,
respectively. As expected, the temperature of the cold dust
is very similar to that of the single blackbody
fit. We also note that the temperature of the warm dust is not
  tightly constrained, as $T_{\rm dust}^w= 67\pm 5\,$K produces 
values of $\chi^2$ similar to $\chi^2_{\rm min}$. Finally, the
goodness of the fit of the two temperature model is only marginally
better than the single temperature fit.
\citet{Clements2010} found higher dust temperatures for ULIRGs
than those of lower luminosity systems, i.e. a median dust temperature
of 42\,K for a sample of ULIRGs, while a median of 35\,K has been
found for normal galaxies \citep{Dunne2001}. The latter is consistent
with the relatively cold temperature found in Mrk~938. We also find
that the mass ratio of the cold to the warm dust component is very
high ($M_{\rm dust}^{\rm c}/M_{\rm dust}^{\rm w}\sim 79$), which
clearly indicates that the cold component is dominant in this
galaxy. The total dust mass is very similar to the value obtained
  with one dust temperature.

In Fig. \ref{fig:SED} (right panel) we have plotted the entire SED
including the torus model of \emph{hot} dust and the best-fit model to
the far-IR data using the combination of two blackbodies
described above. The latter
best-fit temperatures are typical of star forming
regions in the galaxy disk, therefore confirming that the active
nucleus is not responsible for most of the observed dust emission
\citep[as seen in e.g.][]{Hatziminaoglou2010}. Dust in regions of star
formation have peak temperatures between 40 and 70\,K, while radiation
of AGN origin has a characteristic peak temperature within the range
120$-$170\,K \citep[e.g.][]{Perez2001}. This is in agreement with what
we found in Sect.~\ref{sec:MIR_evidence} from the spectral
decomposition, where we concluded that the dusty torus only
contributes up to 2\% of the observed IR emission.

As a comparison, \citet{Ramos2011b} studied the dust distribution of
the Seyfert 2 galaxy NGC~3081 also including \herschel\ PACS and SPIRE
data in their analysis. The nuclear emission of this galaxy was
reproduced by a clumpy torus model of warm dust heated by the AGN. The
circumnuclear emission was fitted by a cold dust component at 28\,K
heated by young stars in the galaxy disk, plus a colder component at
19\,K in the outskirts of the galaxy heated by the interstellar
radiation. Similarly, \citet{Radovich1999} decomposed the SED of the
narrow line galaxy NGC~7582 and found an extranuclear emission
dominated by cold dust at 30\,K, plus an additional very cold
component emitting at 17\,K. We do not detect this very cold dust
component in Mrk~938. This is because we cannot resolve any extended
emission in our data, and although this component is likely to exist
its contribution to the total IR luminosity is negligible.

We can finally estimate the size of the IR emitting region. Following \cite{Klaas2001}, for an optically thin (transparent) blackbody with $\beta=2$, the far-IR emitting component is homogeneously distributed in a ``minimum face-on disk'' of radius $r_\tau$ with the form: 

\begin{equation}
r_\tau({\rm pc}) = \left( \frac {M_{\rm dust}/{\rm M}_\odot}{41.7 \pi
  \tau_{100\mu{\rm m}}} \right) ^{0.5} 
\end{equation}

Assuming an opacity at 100\,\micron\ $\tau_{100\mu{\rm m}}\sim 0.5- 1$, we obtain a minimum radius for the IR emitting region of $r_\tau \sim 500-800\,$pc, in good agreement with our measurements from the mid-IR images in Sect. (\ref{sec:IRregion}).

\section{Summary and conclusions}

LIRGs have emerged as an important cosmological class given that they may play a central role in our understanding of the general evolution of galaxies and black holes. We have studied Mrk~938, a LIRG in the local Universe classified as a Seyfert 2 through optical spectroscopy. 
Table \ref{table:Mrk938properties} summarises the relevant parameters resulting from the present work which, in addition to table~8 in \citet{Schweizer2007}, shows the main properties of Mrk~938. The analysis of the X-ray data favours the presence of a highly obscured AGN, where the SB emission prevails at soft energies and the active nucleus dominates at hard energies. The absorption-corrected hard X-ray luminosity is too high to be originated only in a pure starburst.

\begin{table*}
  
\caption{Summary of the properties of Mrk~938 derived in this paper.}
\label{table:Mrk938properties}

\begin{tabular}{l c c c}        
\hline
\multicolumn{4}{c}{AGN Properties}\\
\hline
Hard X-ray luminosity & $L_{\rm 2-10keV}$& $1.4^{+0.4}_{-0.2}\times 10^{42}\,{\rm erg \, s}^{-1}$ & absorption corrected\\ 
Bolometric luminosity & $L_{\rm bol}$(AGN) &$1.6^{+1.1}_{-0.6}\times 10^{43}\,{\rm erg \, s}^{-1}$ & from X-ray and bolometric correction\\
                      &                   &$2.6^{+2.6}_{-1.3}\times 10^{43}\,{\rm erg \, s}^{-1}$ & from IRS decomposition\\
AGN contribution      & Bolometric    & 2$^{+2}_{-1}$\% &  from IRS decomposition\\
                      & $0.5-2\,$keV  & $\le 5\%$ & from comparison with SFR-due X-ray luminosity \\
                      & $2-10\,$keV   & $\ge 85\%$  & from comparison with SFR-due X-ray luminosity \\
                      & 24\,\micron    & 5 ($\pm3$)\% & from IRS decomposition\\
\hline
\multicolumn{4}{c}{SB Properties}\\
\hline
Star Formation Rate  &  SFR$_{\rm obscured}$ & $42\,{\rm M}_\odot$ yr$^{-1}$ & from MIPS 24\,\micron, Kroupa IMF\\

Size IR emitting region &  $d_{\rm IR}$   & $\sim 1\,$kpc & from IRAC 8\,\micron\ (FWHM) \\
                        &                &  $\le 2\,$kpc & from MIPS 24\,\micron, PACS 70\,\micron\ (FWHM) \\
                        &                  & $1-1.6\,$kpc & from dust mass, assuming $\tau_{100\mu{\rm m}}=0.5-1$\\
Dust Temperature        &    $T_{\rm dust}^{\rm c}$ & 35\,($\pm$4)\,K &  two temperature fit, fixed $\beta=2$\\ 
                        &    $T_{\rm dust}^{\rm w}$ & 63\,($\pm$4)\,K &  ``\\
Dust Mass               &   $M_{\rm dust}$ & $3(\pm1)\times 10^7\,{\rm M}_\odot$ & ``\\
Dust mass Ratio      & $M_{\rm dust}^{\rm c}/M_{\rm dust}^{\rm w}$ & 79 & `` \\          
\hline
\end{tabular}
\end{table*}

We have performed a decomposition of the \spitzer/IRS spectrum and recovered the AGN contribution to the IR luminosity, which is overwhelmed by the intense emission of star formation. The best fit model to the IRS data was achieved with a combination of a type 2 AGN plus a $10^{\rm 12}\,{\rm L}_\odot$ starburst templates. Using the derived fractional contribution of the AGN to the luminosity at 24\,\micron\ we derived an AGN bolometric contribution to the IR luminosity of $\sim$2\%, which is in agreement with previous estimations. This supports the proposed scenario that intense dusty starbursts are responsible for the high IR luminosities in most local LIRGs. The analysis of IR images shows that the IR emission of Mrk~938 originates in a relatively compact region, with a size between 0.5 and 2\,kpc. We have also demonstrated that most of the on-going SFR is produced in an obscured environment as expected for LIRGs.

The wealth of multi-wavelength archival data of this source has been used in conjunction with our own observations to constrain the SED. We have used \herschel\ imaging data for the first time to constrain the cold dust emission with unprecedented accuracy.  We have derived dust temperatures of 63\,K and 35\,K for the warm and cold components, respectively. Using the results of our SED fitting we have derived a cold dust mass of $3\times10^{7}\bf ,{\rm M}_\odot$ and a very high mass ratio of the cold to the warm dust components. This demonstrates that emission of cold dust clearly dominates the IR SED.

\section*{Acknowledgments}
Thanks to F.~Schweizer for kindly providing the optical image of Mrk~938, to J.~Gallimore for providing the MIPS SED data, and to H.~Krimm and W.~Baumgartner for the analysis of the BAT observations. PE, AAH and MPS acknowledge support from the Spanish Plan Nacional de Astronom\'{\i}a y Astrof\'{\i}sica under grant AYA2009-05705-E. MPS acknowledges support from the CSIC under grant JAE-Predoc-2007. AMPG acknowledges support by the Spanish {\it Plan Nacional de Astronom\'ia y Astrof\'isica} under the grant AYA2008-06311-CO2-01. CRA acknowledges financial support from STFC (ST/G001758/1) and from the Spanish Ministry of Science and Innovation (MICINN) through project Consolider-Ingenio 2010 Program grant CSD2006-00070: First Science with the GTC. MP acknowledges Junta de Andaluc\'{\i}a and Spanish Ministry of Science and Innovation through projects PO8-TIC-03531 and AYA2010-15169.

PACS has been developed by a consortium of institutes led by MPE (Germany) and including UVIE (Austria); KU Leuven, CSL, IMEC (Belgium); CEA, LAM (France); MPIA (Germany); INAF-IFSI/OAA/OAP/OAT, LENS, SISSA (Italy); IAC (Spain). This development has been supported by the funding agencies BMVIT (Austria), ESA-PRODEX (Belgium), CEA/CNES (France), DLR (Germany), ASI/INAF (Italy), and CICYT/MCYT (Spain). SPIRE has been developed by a consortium of institutes led by Cardiff University (UK) and including Univ. Lethbridge (Canada); NAOC (China); CEA, LAM (France); IFSI, Univ. Padua (Italy); IAC (Spain); Stockholm Observatory (Sweden); Imperial College London, RAL, UCL-MSSL, UKATC, Univ. Sussex (UK); and Caltech, JPL, NHSC, Univ. Colorado (USA). This development has been supported by national funding agencies: CSA (Canada); NAOC (China); CEA, CNES, CNRS (France); ASI (Italy); MCINN (Spain); SNSB (Sweden); STFC (UK); and NASA (USA). This work is based on observations made with the \spitzer\ Space Telescope, which is operated by the Jet Propulsion Laboratory, California Institute of Technology, under NASA contract 1407.

\bibliographystyle{aa}
\bibliography{Mrk938}

\label{lastpage}

\end{document}